\documentclass[12pt]{iopart}

\usepackage{graphicx}
\usepackage{epsfig}
\usepackage{amssymb}
\usepackage{color}

\newcommand{\be}{\begin{equation}}
\newcommand{\ee}{\end{equation}}
\newcommand{\ba}{\begin{eqnarray}}
\newcommand{\ea}{\end{eqnarray}}

\begin{document}

\title[Shear viscosity in a  superfluid   cold Fermi gas at unitarity]{Shear viscosity in a  superfluid   cold Fermi gas at unitarity}

\author{M. Mannarelli $^1$, C. Manuel $^2$ and  L. Tolos$^{2,3}$}

\address{$^1$ I.N.F.N., Laboratori Nazionali del Gran Sasso, Assergi (AQ), Italy}
\address{$^2$ Instituto de Ciencias del Espacio (IEEC/CSIC).
Campus Universitat Aut\` onoma de Barcelona,
Facultat de Ci\` encies, Torre C5
E-08193 Bellaterra (Barcelona), Spain}
\address{$^3$ Frankfurt Institute for Advanced Studies, Johann Wolfgang Goethe University, Ruth-Moufang-Str. 1
60438 Frankfurt am Main}
\ead{massimo@lngs.infn.it}

\begin{abstract}

We analyze the contributions of Nambu-Goldstone modes to the shear viscosity of a superfluid atomic Fermi gas close to unitarity.  We show that the low temperature experimental values of the shear viscosity to entropy ratio, $\eta/s$, are reproduced considering an effective shear viscosity which takes into account the collisions among Nambu-Goldstone bosons and the finite size of the optical trap. We predict that, for $T \lesssim 0.1 T_F$, $\eta/s$ should correlate with the size of the trap and linearly decrease with decreasing temperature. For $T \gtrsim 0.1 T_F$, we find agreement with the experimental data assuming that the Nambu-Goldstone modes have an anomalous dispersion law.

\end{abstract}

\maketitle

The knowledge of  the transport coefficients of a fluid may  provide, among other things,  a key to understand the microscopic physics of the system \cite{Schafer:2009dj,Giorgini:2008zz}. After the conjecture \cite{hep-th/0104066}, derived by string theory methods, of the existence of a universal bound of the shear viscosity to entropy ratio $\eta/s \geq  \hbar/ ( 4 \pi k_B)$,  special attention has been drawn to those fluids that could saturate the bound,
as it has been claimed that their microscopic degrees of freedom cannot be identified with well-defined quasiparticles.  The fluid produced in heavy-ion collisions seems to have extremely small shear viscosity~\cite{Schafer:2009dj}, close to the universal bound, and almost ideal hydrodynamical flow was observed  in  a dilute Fermi gas of cold atoms at unitarity~\cite{Ohara:2002}, {\it i.e.} with an infinite two-body scattering length.

Experiments with trapped cold atomic Fermi  gases are able to reach the region of infinite scattering length  tuning the interaction between the fermionic atoms by means of  a magnetic-field Feshbach resonance \cite{Ohara:2002,Bourdel:2003zz,Gupta:2003,Regal:2003}. The strength of the interaction between atoms  depends on  the applied  magnetic field and can be  measured  in terms of the $s$-wave scattering length. By varying the magnetic-field controlled interaction,  fermionic pairing is observed to undergo the Bose-Einstein condensate  to Bardeen-Cooper-Schrieffer crossover.  It is a remarkable aspect  of these fermionic  systems that  for any value of the  attractive interaction  they are  superfluid, provided the temperature is sufficiently low.

The experimental setup used to measure  the shear viscosity  consists of a degenerate mixture of an equal number of spin-1/2-up and spin-1/2-down optically trapped fermionic atoms. In particular we shall refer to the results reported in Refs.~\cite{kinast1, kinast3} with  $N \simeq 2.0 \times 10^5$ $^6$Li atoms  in a magnetic field of $B= 840$ G,   close to the $s$-channel Feshbach resonance at $B=834.15$ G \cite{Bartenstein}.   
The unperturbed density in the trap is given by (in units  $\hbar = k_B=1$, hereafter used)
\be
\rho_0({\bf r}) =\frac{(2 m E_F)^{3/2}}{3 \pi^2} \left(1-\sum_{i=1}^3 \frac{r_i^2}{R_i^2}\right)^{3/2} ~~~ R_i=\sqrt{\frac{2 E_F}{m \omega_i^2}} \,,
\ee
where $m$ is the mass of the  $^6$Li atom and $E_F$ is the Fermi energy of $N$ free fermions in a harmonic 
oscillator potential.  In the experimental setup of  Refs.~\cite{kinast1, kinast3},  the transverse frequencies of the trap are   $\omega_x\simeq 2\pi\times 1778\,{\rm s}^{-1}$, 
$\omega_y \simeq 2\pi\times 1617\,{\rm s}^{-1}$
and the net axial frequency is $\omega_z\simeq 2\pi\times76\,{\rm s}^{-1}$, corresponding to a  Fermi temperature   $ T_F \simeq 2.4~10^{-6} {\rm K}  $.  

Values of the viscosity can be extracted at both high and low temperatures by studying either the anisotropic expansion of the  atomic cloud or  the breathing mode damping \cite{Cao:2010wa}. At high $T$ a dimensional analysis, or a much careful study based on a Boltzmann equation, predicts the scaling  $\eta \propto T^{3/2}$ \cite{aboveTc}, in agreement with the experimental results~\cite{Cao:2010wa}. At low $T$, below the
superfluid phase transition,  the theoretical and experimental study of the transport properties of the system is quite challenging. 

 Superfluidity is a phenomenon that occurs after the appearance of a quantum condensate that spontaneously breaks a global symmetry of the system~\cite{IntroSupe}. 
 First discovered in  $^4$He, it was soon realized that the phenomenon occurs in both bosonic and fermionic systems. The property of superfluidity follows from the existence of Nambu-Goldstone modes, hereafter referred to as   phonons, associated to the breaking of
 the particle number conservation law.   In the present  manuscript, we employ  the same  mechanism that explains the behavior of $\eta$ in  $^4$He (in the regime  where phonons dominate)  to the evaluation of the viscosity in the superfluid phase of the cold Fermi atoms in the unitarity limit. 
 
In our analysis we neglect the contribution of  fermionic modes to the shear viscosity.  This approximation should be reliable for very low temperatures, $T \lesssim 0.1 T_F$, where the density of superfluid fermionic modes is exponentially suppressed and the  normal fluid corona around the center of the trap is small. Actually,
 the results of \cite{Guo:2010dc}   show that the fermionic contribution to $\eta$ is above the experimental values for any value of  $T < T_c$.  Since $\eta$ is directly proportional to the typical time for transport of momentum in the direction orthogonal to the flow, $\tau_{\perp}$,   there should be modes with a shorter $\tau_{\perp}$. We assume that at very low temperature this mode is the phonon.

 Let us briefly review some well known facts about  the computation of $\eta$ in superfluid $^4$He. Landau  postulated a Hamiltonian that dictates the dynamics of the phonons \cite{IntroSupe} and at very low temperature Landau and Khalatnikov computed the shear viscosity due to   binary collisions of phonons, finding the scaling
 $\eta \propto 1/T^5$ \cite{Landau}. Experimental measures of $\eta$   showed however a different scaling, close to $ 1/T$  \cite{Whitworth}.
   Then, Maris had the great insight to formulate that the shear viscosity in $^4$He at low temperature was instead dominated by  $1\leftrightarrow 2$ processes (hereafter 3ph) small-angle collisions \cite{Maris}, only possible due to an anomalous dispersion law of these   collective modes. He further realized that the knowledge of the $T$ scaling behavior of $\eta$ could provide information on the momentum behavior of the phonon dispersion law. A detailed  analysis of the experimental results has lead to the conclusion that  a complex hierarchy of time-scales exists and that in order to properly reproduce the experimental results the transition from the hydrodynamic to the ballistic regime of phonons should  be taken into account~\cite{Zadorozhko, Niemetz, Morishita, Eselson}. 
   
In the superfluid phase of the cold Fermi system at unitarity  effective field theory techniques have been applied to construct the Lagrangian of the phonons in an energy and momentum expansion \cite{Son:2005rv}. At the lowest order in the expansion the Lagrangian reads
\be
\label{LO-Lagrangian}
{\cal L}_{\rm LO} = P(X)  , \qquad X = \mu - V({\bf r}) - \partial_0 \phi - \frac{({\bf \nabla} \phi)^2}{2 m} \,,
\ee 
where $P$ is the pressure of a  superfluid at $T=0$, see  {\it e.g.}     \cite{Son:2005rv},  $\phi$ is the phase of the condensate and $V({\bf r})$ is the trapping  potential. For vanishing trapping potential,  the formal expression of  ${\cal L}_{\rm LO}$   can be related by a Legendre transformation to the Hamiltonian proposed by Landau to derive the phonon self-interactions.   The Lagrangian at the next to leading order (NLO) in a momentum expansion has as well been constructed in  \cite{Son:2005rv} and leads to a phonon dispersion law of the form $E_p = c_s p (1 + \gamma p^2)$ where $c_s$ is the speed of sound, 
\be
\label{B-conformal}
 \gamma = - \pi^2  \sqrt{2 \xi}\left(c_1+ \frac{3}{2} c_2\right) \frac{1}{k_F^2}\,,\ee
 where $k_F$ is the Fermi momentum, and $\xi\simeq 0.37$ \cite{Ku,  Haussmann:2007zz, Arnold:2006fr} is the Bertsch number.  The two parameters $c_1$ and $c_2$, related to the momentum dependence of the static density and to the transverse response functions,  can be evaluated  by the $\epsilon$-expansion  technique \cite{Rupak:2008xq},
finding $c_1 \simeq -0.021$, and $c_2/c_1 = {\cal O}(\epsilon^2)$. They can also  be obtained by a fit of  Monte Carlo numerical simulations \cite{Salasnich}, which gives similar results.
The interesting fact of these results is that $\gamma > 0$, and therefore  3ph small-angle collisions are kinematically allowed. The contributions of these processes to the bulk viscosity coefficients and to the thermal conductivity of ultracold atomic  systems have been computed in Ref.~\cite{Escobedo:2009bh} and in Ref.~\cite{Braby:2010ec}, respectively. Nonetheless, the results of Ref.~\cite{Haussmann:2009}, obtained by combining the $\epsilon$-expansion technique with a gaussian expansion around the mean field, suggest that $\gamma < 0$. But, considering gaussian fluctuations close to  unitarity might not be enough to determine the dispersion law of the collective modes, because the procedure seems to be not self-consistent \cite{Diener:2008}. Therefore, in the present manuscript we refrain from considering $\gamma < 0$; we shall  consider  how our results would change in that case in a future work.

The shear viscosity due to phonon binary collisions in the cold Fermi fluid has been computed, finding a scaling $\eta_{\rm 4ph} \propto 1/T^5$ \cite{Rupak:2007vp,Manuel:2011ed}, the same as in $^4$He, which however is not consistent with experimental values of $\eta$. 
This fact motivates the inclusion of 3ph small-angle collisions which are kinematically allowed if the phonon dispersion law is of the form
\be\label{disp} E_p = c_s p (1 + \psi(p))\, ,\ee 
where $\psi(p) $ is positive in a certain range of momenta. We shall assume that $\psi(p) \ll 1$, and   carry  out the computation of the shear viscosity in a perturbative expansion in $\psi(p)$.

For the evaluation of the contribution of 3ph proceses to the shear viscosity, $\eta_{\rm 3ph},$ we consider the phonon transport equation amended by a collision term that takes into account $1 \leftrightarrow 2$ splitting and joining process. In order to simplify the analysis, in this  calculation we neglect the effect of the trapping potential, meaning that  we evaluate the scattering
matrix associated to these processes  from the Lagrangian in Eq.~(\ref{LO-Lagrangian}) taking $V({\bf r}) \equiv 0$. We obtain that  the square of the scattering amplitude reads
\be
|{\cal M}|^2= 4 c_s^4  \frac{\left( 1+ u \right)^2}{\rho}( p p' k' )^2 + {\cal O}(\psi)\,,
\ee
where $\rho$ is the mass density of the system, $u = \frac{\rho}{c_s} \frac{\partial c_s}{\partial \rho}$
and $p, p',k'$ refer to the modulus of the momenta of the three phonons participating in the process. The expression of this scattering amplitude is formally equivalent, as one could naively expect, to the corresponding expression obtained for 3ph interactions in  $^4$He. What changes are the explicit values of $\rho$ and $c_s$ in the two different systems.

We have solved the Boltzmann transport equation using variational methods, and find a solution that fully agrees with that proposed to explain the phonon viscosity due to small-angle collisions  in $^4$He~\cite{Benin}. In particular, we find that the contribution of 3ph processes to the shear viscosity coefficient is given by
\be
\label{3phshear}
\eta_{\rm 3ph} = \left( \frac{2 \pi}{15}\right)^4 \frac{T^8}{c_s^8}  \frac{1}{M} + {\cal O} ( \psi^2) \ ,
\ee
where
\ba
\label{valueM}
M =  \frac{1}{20  T \pi^3}  \frac{   \left( 1+ u \right)^2}{\rho}
\int^* dp' \,dk'\, (p' k' ( p' +k') )^2 |{\cal M}|^2 \nonumber \\
  \left( b(p') + b(k') - b(p'+k') \right)^2 
f^{\rm eq}_{p'+k'}  ( 1+f^{\rm eq}_{p'})(1+  f^{\rm eq}_{k'})\,, \ea
where $f^{\rm eq}_p$ is the equilibrium Bose-Einstein distribution function for the phonon of
momentum $p$, and we have defined $b(p) = p \psi(p)$.  The star on the top of the integral indicates that the integration is constrained
by the energy and momentum conservation laws. From the expression above one clearly sees that
different forms of the phonon dispersion law may produce different $T$ scalings of $\eta_{\rm 3ph}$.

Assuming that  $\psi(p) = \gamma p^2$, where $\gamma >0$ is given by Eq.~(\ref{B-conformal}), one finds $\eta_{\rm 3ph} \propto 1/T^5$, which is the same scaling found considering   binary collisions. This result is in disagreement with the 
experimental data both at low temperature, because   $\eta_{\rm 3ph} $ diverges, and at high temperature, because one finds that $\eta_{\rm 3ph} < 1/ (4 \pi)$ and below the experimental values. This last result  motivated us to consider, as in the $^4$He system, a dispersion law with 
\be
\label{psi} \psi(p) = \gamma p^2 - \delta p^4\,,
\ee 
 with $\delta > 0$, corresponding to phonons with anomalous dispersion. Upon substituting Eq.~(\ref{psi}) in Eq.~(\ref{valueM}) we find that
\be\label{eta-3ph}
\eta_{\rm 3ph} =  \frac{2^6  \pi^7 }{3^4 5^3} \frac{c_s^2 \, \rho}{T (1+u)^2} \frac{1}{\psi^2_{\rm max}\, I(\tilde \beta)} \,,
\ee
where $\psi_{\rm max} = \frac{\gamma^2}{4 \delta}$ and ${\tilde \beta} = \frac{2 \delta}{ \gamma}\frac{T^2}{ c_s^2}$, while
$I(\tilde \beta)$ is, up to some normalization constants, the integral that appears in Eq.~(\ref{valueM}) after rescaling the momenta so as to make the integral over adimensional variables. We then find that in the region, $0.1\,T_F \lesssim T  \lesssim 0.17\,T_F$, $I(\tilde\beta) \approx  2046.4 $, independent of $T$, so that 
 $\eta_{\rm 3ph} \propto  1/T$. For $T \lesssim 0.1\,T_F $ only the linear and cubic terms in momenta in the phonon dispersion law are relevant, and we find that  $\eta_{\rm 3ph} \propto 1/T^5$, in agreement with the result obtained with vanishing $\delta$.

As we shall see,  the shear viscosity in  Eq.~(\ref{eta-3ph}) gives a good agreement with the experimental data in the range $ 0.1\, T_F \lesssim T \lesssim  T_c$, but it still diverges when the temperature is further decreased. However, the measured shear viscosity of the system does not diverge and we shall show that this behavior can be reproduced including   finite size effects,  which start to  play a relevant role at very low temperature. Indeed, we find that at $T \sim 0.1 T_F$  the phonon mean free path, $\ell_{\rm ph}$,  becomes larger than the size of the system. Then, phonons are ballistic and cannot be described by means of hydrodynamics. However,  it is still possible to define an effective ballistic viscosity, which characterizes the damping of the fluid oscillations, in the transition regime. The existence of the effective viscosity is linked to the presence of a boundary and to the fact that quasiparticles can scatter at the boundary. In experiments with $^4$He it has been shown that the effective shear viscosity that takes into account the ballistic phonons  
\be\label{ball}
\eta_{\rm ball} = \frac{1}{5}  \rho_{\rm ph} c_s a \,,
\ee
leads to excellent agreement with the experimental data \cite{Zadorozhko, Niemetz, Morishita, Eselson}. Here,  $\rho_{\rm ph} = \frac{2 \pi^2 T^4}{45 c_s^5}$ is the density of the normal fluid component and $a$  is the  typical length scale of the  system, which correlates with the size of the device used to  measure the shear viscosity, see {\it e.g} \cite{Zadorozhko}. Analogous equations can be derived for other fluids, see e.g. \cite{Hadjiconstantinou}.
In the following we shall assume that the same expression for the effective shear viscosity of the superfluid phonons   given  in Eq.~(\ref{ball}) holds for the phonons of the cold Fermi gas.
 
In the cold Fermi gas  the relevant dimension is the shortest dimension of the trap and  since $R_x$ is the shortest radius, we assume that  $a < R_x$. The actual value of $a$ depends on the extension of the superfluid region at the trap  center  and on the properties of the boundary between the superfluid region and the corona of unpaired fermions. We postpone the evaluation of $a$ and  the detailed treatment of the trap to future work and consider that the system consists of an homogenous superfluid  up to some distance $a$ which is independent of the temperature. Then we  determine the numerical value of $a$ from the experimental data. Since $a$ is related to the extension of the superfluid region, it should decrease with increasing temperature. However, in the region $T < 0.1 T_F$, where phonons are ballistic, the extension of the superfluid region is not strongly dependent on the temperature, see e.g. \cite{Haussmann:2008}.

Assuming that the 3ph collisions and the interactions with the boundary can be treated as independent processes, the effective shear viscosity due to phonons is given by \cite{Zadorozhko} 
\be
\eta_{\rm eff} = \left( \eta_{\rm  3ph}^{-1} + \eta_{\rm ball}^{-1}\right)^{-1} \,.
\label{eta}
\ee
 
The effective shear viscosity  depends on two unknown quantities, $\psi_{\rm max}$ and $a$, which we shall use as fitting parameters. In principle, the 4ph processes discussed in \cite{Rupak:2007vp,Manuel:2011ed} should be included in the definition of the effective shear viscosity coefficient. However, $\eta_{\rm 4ph}$ has only been evaluated  using the LO dispersion law and therefore cannot be consistently included in the present analysis.

\begin{figure}
\begin{center}
\includegraphics[width=0.7\textwidth]{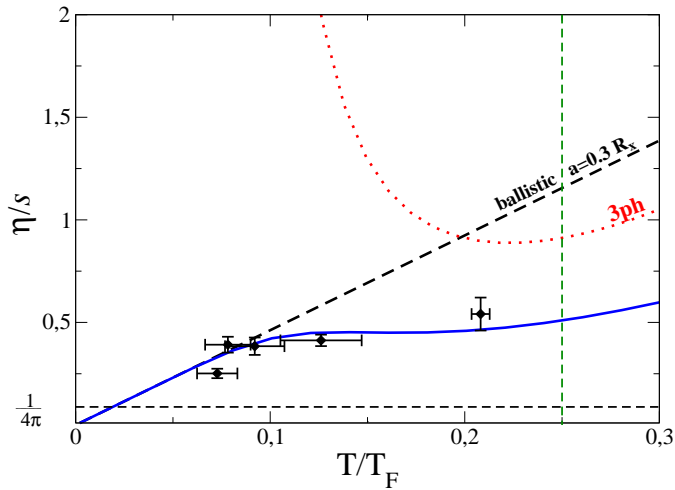}
\caption{\it  Plot of  $\eta/s$ in units of  $\hbar/k_B$ as a function of $T/T_F$. We report all the various contributions separately as well as the effective  shear viscosity obtained with Eq.~(\ref{eta}).  The interaction of the phonons with the boundary (dashed black line)  is obtained from Eq.~(\ref{ball}) assuming $a=0.3\,R_x$; the contribution of the 3-ph  process (dotted  red line) is obtained from Eq.~(\ref{eta-3ph}) considering  $\psi_{\rm max} =0.3$.    
The experimental values and error bars were taken from \cite{Cao:2010wa}. 
The blue solid  line corresponds to the effective shear viscosity obtained by Eq.~(\ref{eta}).
 The green vertical dashed   line approximately corresponds to the transition temperature between the normal phase and the superfluid phase \cite{kinast3}. The actual critical temperature may be lower, $T_c \simeq 0.21 T_F$ as reported in \cite{Haussmann:2008}.  The horizontal  black dashed line corresponds to the universal bound \cite{hep-th/0104066}. } \label{fig:eta-s}
\end{center}
\end{figure}

Assuming that the entropy is also dominated by  phonons, we plot in Fig.~\ref{fig:eta-s} the values of $\eta/s$,  together with the partial contributions $\eta_{\rm 3ph}/s$ (dotted red line) and $\eta_{\rm ball}/s$ (black dashed  line).
We present results obtained with $a=0.3 R_x$ and  $\psi_{max}=0.3$, meaning that $\delta \simeq 0.8\, \gamma^2$ 
(taking $a=(0.2 -0.4) R_x$ and $\psi_{max} = 0.28 - 0.4 $ still gives good agreement with the experimental data below $T_c$; we shall present a more detailed numerical analysis elsewhere).  The green vertical dashed   line approximately corresponds to the transition temperature between the normal phase and the superfluid phase, $T_c \simeq (0.21 -0.25) T_F$. It has to be understood that  the various plots of the phonon contributions  are meaningful only for values of $T  < T_c$, because  phonons do not exist in the normal phase. The   blue solid line corresponds to the effective shear viscosity, $\eta_{\rm eff}$, and is obtained by Eq.~(\ref{eta}).  The agreement of this result with  the experimental data below $T_c $ is remarkable, given that our calculation depends in a non-trivial way on only two parameters. Moreover, notice that the blue solid line  in the range $0.1\,T_F < T < T_c$  is almost independent of the temperature, and this behavior is due to the presence of a minimum of $\eta_{\rm 3ph}/s$ at $T\simeq 0.2\,T_F$, which can be traced back to the $\psi(p)$ correction, given in Eq.~(\ref{psi}), to the phonon dispersion law in Eq.~(\ref{disp}).

The fact that in the range  $0.1 \,T_F \lesssim T \lesssim 0.2\, T_F$  the  blue solid line is close to the experimental values suggests that  phonons  have an anomalous dispersion law. Indeed, assuming that the fermionic processes and the bosonic processes are independent, the process with the smaller value of $\tau_\perp$ (the typical time of transfer of momentum in the direction orthogonal to the flow)  will dominate. Then,  considering that  $\eta \propto \tau_\perp$, it follows that  the experimental values of the shear viscosity represent a lower bound for  the shear viscosity contribution of each component. If the $\delta$$p^4$ correction, given in Eq.~(\ref{psi}), is not  included in the evaluation of $\eta_{\rm eff}$, we find for $T \simeq 0.2 T_F$ values below the experimental ones and below the universal bound $1/(4 \pi)$. Therefore, our simplified model suggests that phonons should have an anomalous dispersion law. The caveat is that with increasing temperature the reduction of the superfluid region and the modification of the matter distribution inside the trap should  be properly included in the evaluation of the shear viscosity coefficient. 
For $T \simeq 0.1 \,T_F$ the largest part of the system is superfluid  \cite{Haussmann:2008} and the matter distribution is not strongly changed with respect to $T=0$, thus we expect the contribution of phonons to be  dominant. But, with increasing temperature  the extension of the corona increases  \cite{Haussmann:2008}, and the fermionic contribution to the shear viscosity should not be negligible, although it is not clear  at which temperature the fermionic and bosonic contributions should become comparable. 
Moreover, it is not clear whether  bosons and  fermions can be treated as independent degrees of freedom, because some relaxation process between phonons and fermions might become important  for $T > 0.1\, T_F$. The proper way to take into account both degrees of freedom is to solve the Boltzmann equation which takes into account the distribution of  fermions and phonons in the trap.  We leave this complicated analysis to future work.

A remarkable result of our analysis is that $\eta/s$ should strongly depend on the trap size. If experiments are conducted increasing (reducing) the size of the trap, but keeping $E_F$ constant, we predict that $\eta/s$  at low temperatures should increase (decrease). The effect should be  sizable at any temperature below $T_c$, and more prominent at $T\lesssim 0.1\,T_F$, where it is independent of the detailed form of the phonon dispersion law. Indeed, at such low temperatures the correction, $\psi(p)$, to the phonon dispersion law is completely negligible and $\eta_{\rm eff}$ is dominated by the ballistic effective viscosity, because the mean free path of phonons is larger than the trap size. 
The experimental investigation  of the ultralow temperature regime would also clarify whether the decrease of $\eta/s$ with decreasing temperature is due to ballistic phonons or to effects associated to noncondensed pairs~\cite{Guo:2010dc}. Indeed, in Ref.~\cite{Guo:2010dc} it was found that $\eta/s$ mildly decreases with decreasing temperature,  while we predict that $\eta/s \propto T$. Moreover, the fermionic contribution  should not scale with the size of the experimental system, while we find that $\eta/s \propto R_x$. Whether  the effective description of the viscosity proposed in the present paper remains valid at such a low temperature or whether some unknown mechanism  prevents $\eta/s$ to vanish with decreasing temperature deserves further investigation.

  It is important to note that the modified dispersion law of phonons, Eq.~(\ref{disp}),  influences all the thermodynamic properties of the system at low $T$. These include the pressure, the entropy or the specific heat, that would get extra $T$ corrections which also
  depend on the values $\gamma$ and $\delta$, if a phonon dispersion law of the form suggested in Eq.~(\ref{psi}) is used. Low thermal corrections to the speed of sound and its attenuation factor might also be sensitive to the higher order corrections to the phonon dispersion law. It was actually the precise
  measurement of the specific heat of the low $T$ regime of superfluid $^4$He \cite{Philips}  that could confirm the form of the phonon dispersion law necessary to explain the value of shear viscosity of that system. It would thus be interesting to have precise measurement of different low $T$ thermodynamical
  variables to confirm the form of the phonon dispersion law that we suggested in this manuscript.

\ack
We thank T.~Schafer for discussion and J.E.~Thomas and C.~Cao  for providing us with the experimental data points shown in Fig.~\ref{fig:eta-s}. M.M. thanks W.~Zwerger for discussion and suggestions. This research was in part supported from Ministerio de
Ciencia e Innovaci\'on under contract FPA2010-16963. 
L.T. acknowledges support from Ramon y Cajal
Research Programme, and from FP7-PEOPLE-2011-CIG under contract
PCIG09-GA-2011-291679

\end{document}